\documentclass{Interspeech}

\usepackage{subcaption}
\usepackage{multirow}
\usepackage{color}
\usepackage[table]{xcolor}
\usepackage{etoolbox}

\definecolor{Gray}{gray}{0.9}

\DeclareMathOperator{\cossim}{sim}
\DeclareMathOperator{\softmax}{softmax}
\DeclareMathOperator{\sample}{sample}
\DeclareMathOperator*{\topk}{top}

\newcommand{\Qpos}{\boldsymbol{Q}^{\prime}}
\newcommand{\qpos}{\boldsymbol{q}^{\prime}}
\newcommand{\Qref}{\hat{\boldsymbol{Q}}}

\usepackage{hyperref}

\interspeechcameraready

\title{SSPS: Self-Supervised Positive Sampling for Robust \\Self-Supervised Speaker Verification}

\author[affiliation={}]{Theo}{Lepage}
\author[affiliation={}]{Reda}{Dehak}

\affiliation[nocounter]{}{EPITA Research Laboratory (LRE)}{France}
\email{theo.lepage@epita.fr, reda.dehak@epita.fr}

\usepackage{comment}

\begin{document}

\maketitle

\begin{abstract}
    

    Self-Supervised Learning (SSL) has led to considerable progress in Speaker Verification (SV). The standard framework uses same-utterance positive sampling and data-augmentation to generate anchor-positive pairs of the same speaker. This is a major limitation, as this strategy primarily encodes channel information from the recording condition, shared by the anchor and positive. We propose a new positive sampling technique to address this bottleneck: Self-Supervised Positive Sampling (SSPS). For a given anchor, SSPS aims to find an appropriate positive, i.e., of the same speaker identity but a different recording condition, in the latent space using clustering assignments and a memory queue of positive embeddings. SSPS improves SV performance for both SimCLR and DINO, reaching 2.57\% and 2.53\% EER, outperforming SOTA SSL methods on VoxCeleb1-O. In particular, SimCLR-SSPS achieves a 58\% EER reduction by lowering intra-speaker variance, providing comparable performance to DINO-SSPS.
\end{abstract}

\section{Introduction}
\label{sec:introduction}

\newcommand\blfootnote[1]{%
  \begingroup
  \renewcommand\thefootnote{}\footnote{#1}%
  \addtocounter{footnote}{-1}%
  \endgroup
}

\blfootnote{Code: \url{https://github.com/theolepage/sslsv}}

The main application of speaker recognition is Speaker Verification (SV), which determines whether a given speech sample matches a claimed identity. SV systems aim to define a representation space that minimizes inter-speaker similarities and maximizes intra-speaker similarities while ensuring robustness against extrinsic variabilities (e.g., environmental noise, channel, and mismatching recording devices). Deep learning has significantly advanced the field, surpassing traditional approaches such as i-vectors~\cite{dehak2011IVector} with models such as x-vectors~\cite{snyder2018XVectors}, ResNet~\cite{chung2020DefenceMetricLearningSR}, and ECAPA-TDNN~\cite{desplanques2020ECAPATDNN} architectures. These methods learn to associate speech samples with their speaker identities in a supervised fashion on large-scale labeled datasets~\cite{chung2018VoxCeleb2}. The effectiveness of deep learning models improves with larger training datasets. However, this reliance on extensive labeled data poses a significant challenge, as high-quality annotated speech samples are scarce and expensive.

Self-Supervised Learning (SSL) has emerged as a promising approach to overcome this limitation by deriving informative representations directly from the input data. Taking advantage of the vast availability of unlabeled speech, SSL enhances model scalability and reduces the reliance on annotated datasets. Several SSL frameworks have been developed around the joint-embedding architecture, where an anchor and a positive are derived from different views of the same input data, preserving the underlying high-level information. Methods based on contrastive learning, such as SimCLR~\cite{chen2020SimCLR} and MoCo~\cite{he2020MoCo}, seek to maximize the similarity within positive pairs while minimizing the similarity within negative pairs, which are sampled from the batch or a larger memory queue. Self-distillation, such as DINO~\cite{caron2021DINO}, leverages knowledge distillation and the student-teacher paradigm, where a student model is trained to match the teacher model's output. For SV, methods predominantly rely on contrastive learning~\cite{huh2020APAAT,zhang2021SimCLR,xia2021MoCo}, and self-distillation~\cite{cho2022DINO,heo2022DINOCurriculum,zhang2022C3DINO,chen2023RDINO}. These approaches rely on the assumption that the anchor and the positive are from the same speaker identity since both frames are derived from the same utterance. Thus, data-augmentation is fundamental to avoid encoding channel information, shared between the two segments.

However, data-augmentation alone cannot mitigate the impact of SSL same-utterance positive sampling, as it introduces channel characteristics coming from the recording condition into speaker representations, increasing intra-speaker variance. To address this, several methods have been proposed: AP+AAT~\cite{huh2020APAAT} employs an adversarial loss to discourage the model from encoding channel information; i-mix~\cite{kang2022LMix} applies data-driven augmentation by interpolating training utterances to emphasize key distinguishing features; DPP~\cite{tao2023DPP} finds diverse positives by relying on speech and face data; CA-DINO~\cite{han2024CADINO} performs clustering to select positives from the anchor class. Additionally, alternative SSL positive sampling strategies have been explored in computer vision, notably NNCLR~\cite{dwibedi2021NNCLR} and GPS-SSL~\cite{feizi2024GPSSSL}, which identify positives in the latent space using nearest neighbor search.

This paper presents a novel positive sampling strategy for SSL, termed \textbf{\underline{S}elf-\underline{S}upervised \underline{P}ositive \underline{S}ampling (SSPS)}. Rather than selecting a positive sample from the same utterance as the anchor, SSPS identifies a pseudo-positive from a distinct utterance by leveraging the knowledge progressively acquired through SSL. After several epochs of conventional SSL training, it is assumed that samples from the same speaker, recorded under different conditions, will have representations close to the anchor representation. This approach enables learning more robust speaker representations by matching various recording conditions with the same speaker identity. SSPS effectively enhances SV performance across major SSL frameworks by reducing intra-speaker variance. For a detailed analysis and additional results, see the extended version \cite{lepage2025SSLSVBootstrappedPositiveSampling_submitted}.

SSPS is presented in Section~\ref{sec:methods}, following a brief description of the SimCLR and DINO SSL frameworks. The experimental setup is described in Section~\ref{sec:setup}. The impact of SSL positive sampling is first highlighted, followed by a study of SSPS hyperparameters, a comparison of SSPS performance on SV with SOTA methods, and a visualization of speaker representations in Section~\ref{sec:results}. Finally, the article concludes in Section~\ref{sec:conclusions}.

\begin{figure*}
  \centering
  \includegraphics[width=\textwidth]{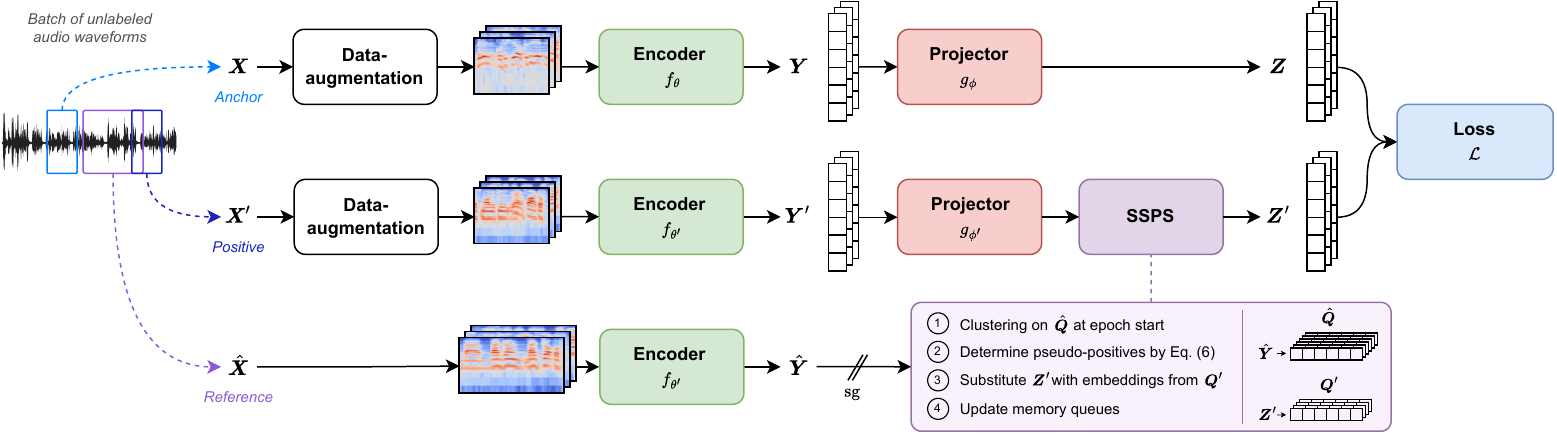}
  \caption{SSL training framework for SV with Self-Supervised Positive Sampling (SSPS).}
  \label{fig:SSPS_SSL_SV_training_framework}
\end{figure*}
\section{Methods}
\label{sec:methods}

\subsection{SSL frameworks}
\label{sec:sslsv}
The self-supervised training framework uses a joint-embedding architecture to create a pair of embeddings from an unlabeled audio sample. Consider a training set of size $N$, with indices denoted by $\mathcal{I} \equiv\{1, \ldots, N\}$. At each iteration, $B$ utterances are selected with $\mathcal{B} \subseteq \mathcal{I}$ the batch indices. From a given utterance $u_i \in \{u_i\}_{i \in \mathcal{B}}$, two segments, $\boldsymbol{x}_i$ (anchor) and $\boldsymbol{x}_i^{\prime}$ (positive), are randomly extracted. Then, random data-augmentation is applied and their mel-scaled spectrograms are used as input features.
The architecture revolves around two branches: (1) an encoder $f_{\theta}$ followed by a projector $g_{\phi}$; (2) an encoder $f_{\theta^{\prime}}$ followed by a projector $g_{\phi^{\prime}}$. Encoders $f_{\theta}$ and $f_{\theta^{\prime}}$ map $\boldsymbol{x}_i$ and $\boldsymbol{x}_i^{\prime}$ to representations $\boldsymbol{y}_i$ and $\boldsymbol{y}_i^{\prime}$ with dimension $D_{\text{repr}}$. Projectors $g_{\phi}$ and $g_{\phi^{\prime}}$ transform $\boldsymbol{y}_i$ and $\boldsymbol{y}_i^{\prime}$ to embeddings $\boldsymbol{z}_i$ and $\boldsymbol{z}_i^{\prime}$ with dimension $D_{\text{emb}}$. Representations are employed for speaker verification, while embeddings are used to compute the loss $\mathcal{L}$. Batches are denoted as $\boldsymbol{X}=\{\boldsymbol{x}_i\}_{i \in \mathcal{B}}$, $\boldsymbol{Y}=\{\boldsymbol{y}_i\}_{i \in \mathcal{B}}$, and $\boldsymbol{Z}=\{\boldsymbol{z}_i\}_{i \in \mathcal{B}}$, with their corresponding counterparts of the other branch denoted as $\boldsymbol{X}^{\prime}$, $\boldsymbol{Y}^{\prime}$, and $\boldsymbol{Z}^{\prime}$, respectively.
By default, SSL frameworks rely on the \textit{symmetrical} joint-embedding architecture where the weights are identical across branches (e.g., SimCLR). When employing the \textit{asymmetrical} version, one branch is designated as the \textit{student} and the other as the \textit{teacher} (e.g., DINO). In this case, the gradient is not computed for the teacher since its weights are updated using an Exponential Moving Average (EMA) of student weights with $m \in [0, 1)$ the momentum coefficient.

\subsubsection{SimCLR}
SimCLR \cite{chen2020SimCLR} is based on contrastive learning as it aims at maximizing the similarity within anchor-positive pairs while maximizing the distance between anchor-negative pairs. Positives are derived from the same utterances as their anchor (same speaker throughout an utterance), and negatives are sampled from the current batch by assuming that negatives will belong to a different speaker identity. $\mathcal{L}_{\text {SimCLR}}$ is defined as:
\begin{equation}
    \label{eq:simclr_loss}
    \mathcal{L}_{\text{SimCLR}} = - \frac{1}{B} \sum_{i \in \mathcal{B}} \log \frac{\exp\left(\cossim\left(\boldsymbol{z}_i, \boldsymbol{z}_i^{\prime}\right) / \tau\right)}{\sum_{j \in \mathcal{B}} \exp\left(\cossim\left(\boldsymbol{z}_i, \boldsymbol{z}_j^{\prime}\right) / \tau\right)},
\end{equation}
where $\cossim(\boldsymbol{a}, \boldsymbol{b})$ represents the cosine similarity between $\boldsymbol{a}$ and $\boldsymbol{b}$, and $\tau$ is a temperature hyperparameter.

\subsubsection{DINO}
DINO \cite{caron2021DINO} adopts a self-distillation training framework in which a \textit{student} is trained to predict the outputs of a \textit{teacher}. The teacher's weights are updated using an EMA of the student's weights. A larger set of augmented utterances of different lengths is considered, resulting in four short (\textit{local}) and two long (\textit{global}) segments. All inputs are processed by the student, but only global views are processed by the teacher. The student and teacher projectors output an embedding normalized with a temperature-softmax. \textit{Centering} and \textit{sharpening} are applied to the teacher embeddings to prevent one dimension from prevailing while discouraging collapse to the uniform distribution. $\mathcal{L}_{\text{DINO}}$ is defined as:
\begin{align}
    \label{eq:dino_loss}
    \mathcal{L}_{\text{DINO}} =
        \frac{1}{B} \sum_{i \in \mathcal{B}}
        \sum_{t=1}^{2} \sum_{\substack{s=1 \\ s \neq t}}^{2+4}
        H\bigg(
            \frac{\boldsymbol{z}_{i,t}^{\prime} - \boldsymbol{c}}{\tau_{\text{t}}},
            \frac{\boldsymbol{z}_{i,s}}{\tau_{\text{s}}}
        \bigg),
\end{align}
where $H\left(\boldsymbol{a}, \boldsymbol{b}\right)=- \softmax(\boldsymbol{a}) \log \left( \softmax(\boldsymbol{b}) \right)$, $\boldsymbol{z}_{i,t}^{\prime}$ and $\boldsymbol{z}_{i,s}$ are the $t$-th teacher and $s$-th student embeddings of sample $i$, $\tau_{\text{t}}$ is the temperature for the teacher, $\tau_{\text{s}}$ is the temperature for the student, and $\boldsymbol{c}$ is a running mean on the teacher outputs.

\subsection{Self-Supervised Positive Sampling (SSPS)}
\label{sec:ssps}
The performance of SSL frameworks primarily depends on how anchor-positive pairs are defined, as it helps model the distribution of each class \cite{balestriero2023CookbookSSL}. SSL commonly relies on data-augmentation to generate a \textit{positive} for a given \textit{anchor}. However, standard data-augmentation techniques may not be sufficient to represent the diversity of acoustic conditions among samples from the same speaker. Thus, SSL models trained for SV are prone to encoding channel-related information because anchor-positive pairs are derived from the same utterances. Self-Supervised Positive Sampling (SSPS) is proposed to sample positives \emph{from different recording conditions} of the same speaker. It is assumed that SSL same-utterance positive sampling group utterances of the same recordings, with similar channel characteristics, before modeling speaker identities. This implies that the latent space is organized by groups of utterances of the same recording and then by subgroups of utterances of the same speaker. Given a training utterance $u_i$ ($i \in \mathcal{B}$), from which the anchor is sampled, let $pos(i)$ denote the index of an utterance $u_{pos(i)}$ used to sample the positive. While standard SSL approaches create the positive from the same utterance (i.e., $pos(i)=i$), SSPS determines a \textit{pseudo-positive} from a different utterance (i.e., $pos(i) \neq i$) in the latent space based on clustering assignments, as detailed in the following. The SSPS training framework is depicted in Figure~\ref{fig:SSPS_SSL_SV_training_framework}.

\subsubsection{Framework}
To capture the unaltered audio patterns, SSPS introduces a \textit{reference} segment $\hat{\boldsymbol{x}_i}$ sampled from $u_i$ using a longer audio segment and no data-augmentation. Additionally, two memory queues are implemented in the framework: $\Qref$ with size $(|\Qref|, D_{\text{repr}})$ for storing reference representations $\{\hat{\boldsymbol{y}}_i\}_{i \in \mathcal{I}}$, and $\Qpos$ with size $(|\Qpos|, D_{\text{emb}})$ for storing positive embeddings $\{\boldsymbol{z}^{\prime}_i\}_{i \in \mathcal{I}}$. After a pre-defined number of standard SSL training epochs, SSPS is enabled and pseudo-positive embeddings are sampled from $\Qpos$ such that $\boldsymbol{z}_i^{\prime}$ is replaced by $\qpos_{pos(i)}$ in the previously defined SSL objective functions.


\subsubsection{Pseudo-positives sampling}
At the beginning of each SSPS epoch, k-means clustering is performed on reference representations in $\Qref$ to group utterances into $K$ clusters, allowing the assignments to be progressively refined as SSL representations improve. $c_i$ denotes the cluster index of the $i$-th utterance, and $\boldsymbol{m}_k$ represents the centroid for the $k$-th cluster. The proposed method considers the following techniques to determine the cluster $\hat{c}_i$ from which to sample a pseudo-positive.
\begin{itemize}
    \item \textbf{Same-cluster sampling.}
    Utterances from the anchor cluster can be considered as pseudo-positives if $K$ tends to the number of speaker identities in the train set, similarly to CA-DINO \cite{han2024CADINO}, such that:
    \begin{equation}
        \hat{c}_i = c_i.
    \end{equation}
    
    \item \textbf{Neighboring-clusters sampling.}
    According to the assumption that channel-related information is modeled before speaker-related information, utterances from neighboring clusters can also be considered as pseudo-positives when selecting a larger value for $K$, such that:
    \begin{equation}
        \hat{c}_i = \sample \left( \mathcal{C}_{c_i} \right),
    \end{equation}
    where $\sample(S)$ is a uniform random selection from $S$, and $\mathcal{C}_k$ consists of the $M$ nearest clusters to the $k$-th cluster:
    \begin{equation}
        \label{eq:C}
        \mathcal{C}_k \triangleq \mathop{\topk M}\limits_{j \neq k} \big( \{ \cossim \left( \boldsymbol{m}_k, \boldsymbol{m}_j \right), \forall j \in [1, K] \} \big),
    \end{equation}
    where $\topk M \left(S\right)$ returns the indices of the largest $M$ values from $S$ in descending order.
\end{itemize}

\noindent
SSPS selects a pseudo-positive for the $i$-th sample according to $\hat{c}_i$, such that:
\begin{equation}
    \label{eq:pos_i}
    pos(i) = \sample \left( \mathcal{S}_{\hat{c}_i} \right),
\end{equation}
where $\mathcal{S}_c \triangleq \{ i \in \mathcal{I} \text{ s.t. } c_i = c \}$ corresponds to the training sample indices from a given cluster. Note that it falls back to default SSL positive sampling if $\qpos_{pos(i)}$ is not present in $\Qpos$.
\section{Experimental setup}
\label{sec:setup}

\subsection{Datasets and feature extraction}

Models are trained on VoxCeleb2 \cite{chung2018VoxCeleb2} \textit{dev} set, consisting of 1,092,009 utterances distributed among 145,569 recordings from 5,994 speakers. Speaker labels are discarded for the SSL training. The evaluation is conducted on VoxCeleb1 \cite{nagrani2017VoxCeleb} \textit{original} test set. Input features are 40-dimensional log-mel spectrograms extracted with the torchaudio library, using a Hamming window length of 25 ms and a frame-shift of 10 ms. Data-augmentation, including reverberation and background noises, is applied with the Simulated RIR Database \cite{ko2017StudyReverberantSpeechRobustSR} and the MUSAN corpus \cite{snyder2015MUSAN}.

\subsection{SSL frameworks}

The encoder $f$ is either based on Fast ResNet-34 \cite{chung2020DefenceMetricLearningSR} or ECAPA-TDNN (C=1024) \cite{desplanques2020ECAPATDNN} for preliminary and final results, respectively. Output representations have a dimension of $D_{\text{repr}}=512$. The supervised baseline corresponds to a model trained using the AAM-Softmax loss with a scale of 30 and a margin of 0.2. The implementation is based on PyTorch, and the trainings are conducted on 2 $\times$ and 4 $\times$ NVIDIA Tesla A100 80 GB.

\subsubsection{SimCLR}
The duration of audio segments is \SI{2}{\second}. The projector $g$ is discarded as it degrades the performance \cite{lepage2024AdditiveMarginSSLSV}. The loss temperature is $\tau=0.03$. The model is trained during 100 epochs with Adam using a batch size set to 256, and a learning rate of 0.001 which is reduced by 5\% every 5 epochs. 

\subsubsection{DINO}
Local and global segments are four \SI{2}{\second} and two \SI{4}{\second} audio chunks, respectively. The projector $g$ consists of an MLP composed of three linear layers and a final weight-normalized linear layer. The hidden dimensions are set to 2048, 2048, and 256. The last layer maps the $l_2$-normalized embeddings to $D_{\text{emb}}=65,536$ units. The student temperature is $\tau_{\text{s}}=0.1$ and the teacher temperature is $\tau_{\text{t}}=0.04$. For the EMA update, $m$ goes from 0.996 to 1.0 with a cosine scheduler.  The model is trained during 80 epochs with SGD, a weight decay of $5e^{-5}$, a batch size of 128, and a learning rate linearly increased to 0.2 during a 10-epochs warm-up before applying a cosine scheduler.

\subsection{SSPS}

Different positive samplings are compared: `SSL', `SSPS', and `Supervised' (uses train set labels to define anchor-positive pairs). Models are obtained by resuming the SSL training with the corresponding positive sampling for 20 epochs. The duration of the reference frame is \SI{4}{\second}. Memory queue lengths are set to $|\Qref|=N$ and $|\Qpos|=K$. K-means is performed using a PyTorch GPU implementation for 10 iterations.

\subsection{Evaluation protocol}

The scoring of each test trial is the cosine similarity of $l_2$-normalized representation pairs, derived from the full-length utterances. The performance is reported in terms of EER and minDCF with $P_{target} = 0.01$, $C_{\text{miss}}=1$, and $C_{\text{fa}}=1$.

\section{Results}
\label{sec:results}

\subsection{Effect of SSL positive sampling on SV performance}

\begin{table}[t]
  \footnotesize
  \caption{SV performance with SSL and Supervised positive sampling using SimCLR and DINO frameworks (ECAPA-TDNN).}
  \label{tab:pos_sampling_bottleneck}
  \centering
  \begin{tabular}{llS[table-format=1.2]S[table-format=1.4]}
    \toprule    
    \textbf{Method} & \textbf{Pos. sampling} & \textbf{EER (\%)} & \textbf{$\text{minDCF}_\text{0.01}$} \\
    \midrule

    \multirow{2}{*}{SimCLR} & SSL & 6.30 & 0.5286 \\
     & Supervised & \bfseries 1.72 & \bfseries 0.2395 \\
    \midrule

    \multirow{2}{*}{DINO} & SSL & 3.07 & 0.3616 \\
     & Supervised & \bfseries 2.36 & \bfseries 0.2712 \\
    \midrule

    \rowcolor{Gray} Supervised & & 1.34 & 0.1521 \\
    \bottomrule
  \end{tabular}
\end{table}

SimCLR, DINO, and the supervised baseline achieve 6.30\%, 3.07\%, and 1.34\% EER on VoxCeleb1-O, respectively, as shown in Table~\ref{tab:pos_sampling_bottleneck}. The performance of SSL methods is significantly improved when using the train set labels to generate anchor-positive pairs from different recordings of the same speaker with distinct channel characteristics. This supervised positive sampling reduces the EER by $\sim$73\% for SimCLR and $\sim$23\% for DINO, highlighting the negative impact of SSL same-utterance positive sampling.

\subsection{Selection of SSPS hyperparameters}

\begin{table}[t]
  \footnotesize
  \caption{Effect of SSPS hyper-parameters ($K$, $M$) on SV performance using SimCLR (Fast ResNet-34).}
  \label{tab:hyperparams}
  \centering
  \begin{tabular}{lccS[table-format=1.2]S[table-format=1.4]}
    \toprule    
    \textbf{Pos. sampling} & $K$ & $M$ & \textbf{EER (\%)} & \textbf{$\text{minDCF}_\text{0.01}$} \\
    \midrule
    SSL & \multicolumn{1}{c}{} & \multicolumn{1}{c}{} & 9.41 & 0.6378 \\
    \midrule
    \multirow{10}{*}{SSPS} & 6,000 & 0 & 6.63 & 0.5493 \\
    \cmidrule(r){2-5} & 10,000 & 0 & 6.82 & 0.5629 \\
    \cmidrule(r){2-5} & \multirow{3}{*}{25,000} & 0 & 7.30 & 0.5805 \\
    & & 1 & 5.80 & \bfseries 0.5250 \\
    & & 2 & \bfseries 5.73 & 0.5258 \\
    \cmidrule(r){2-5} & \multirow{3}{*}{150,000} & 0 & 8.29 & 0.6170 \\
    & & 1 & 7.54 & 0.5923 \\
    & & 2 & 7.13 & 0.5711 \\
    \midrule
    \rowcolor{Gray} Supervised & & & 3.93 & 0.3900 \\
    \bottomrule
  \end{tabular}
\end{table}

To select SSPS hyperparameters, the preliminary results using the Fast ResNet-34 encoder are reported with different values of $K$ and $M$ in Table~\ref{tab:hyperparams}. The systems are compared against the SSL same-utterance positive sampling baseline, achieving 9.41\% EER on VoxCeleb1-O. As expected, sampling from the anchor cluster ($M=0$) and using a number of classes close to the number of speaker identities in the train set ($K=6{,}000$) reduces the EER to 6.63\%. Sampling from a neighboring cluster ($M=1$) and using $K=25{,}000$ further reduces the EER to 5.80\% and represents the best system for minimizing the minDCF. This value of $K$, smaller than the total number of recordings within the train set (i.e., $\sim$150,000), suggests that some recordings are already grouped in the latent space. These results show that sampling positives from a cluster close to their anchor cluster generates appropriate and diverse anchor-positive pairs, which effectively improve SV performance.

\subsection{Final evaluation and comparison to other methods}

\begin{table}[t]
  \footnotesize
  \caption{Evaluation of SSL methods on SV (VoxCeleb1-O). The results for the top rows are drawn from the literature.}
  \label{tab:eval}
  \centering
  \begin{tabular}{lS[table-format=2.2]S[table-format=1.4]}
    \toprule    
    \textbf{Method} & \textbf{EER (\%)} & \textbf{$\text{minDCF}_\text{0.01}$} \\
    \midrule

    \rowcolor{Gray} AP + AAT \cite{huh2020APAAT} & 8.65 & \\
    \rowcolor{Gray} Contrastive + VICReg \cite{lepage2022LabelEfficientSSLSV} & 8.47 & 0.6400 \\
    \rowcolor{Gray} SimCLR + MSE loss \cite{zhang2021SimCLR} & 8.28 & 0.6100 \\
    \rowcolor{Gray} MoCo + ProtoNCE \cite{xia2021MoCo} & 8.23 & 0.5900 \\
    \rowcolor{Gray} CEL \cite{mun2020CEL} & 8.01 & \\
    \rowcolor{Gray} SSReg \cite{sang2022SSReg} & 6.99 & \\
    \rowcolor{Gray} DINO + Cosine loss \cite{han2022DLGLC} & 6.16 & 0.5240 \\
    \rowcolor{Gray} DINO \cite{cho2022DINO} & 4.83 & 0.4630 \\
    \rowcolor{Gray} DINO + Curriculum \cite{heo2022DINOCurriculum} & 4.47 & \\
    \rowcolor{Gray} CA-DINO \cite{han2024CADINO} & 3.59 & 0.3529  \\
    \rowcolor{Gray} RDINO \cite{chen2023RDINO} & 3.29 & \\
    \rowcolor{Gray} MeMo \cite{jin2024WGVKT} & 3.10 & \\
    \rowcolor{Gray} RDINO + W-GVKT \cite{jin2024MeMo} & 2.89 & 0.3330 \\
    \midrule

    SimCLR & 6.30 & 0.5286 \\
    \textbf{SimCLR-SSPS} & \bfseries 2.57 & \bfseries 0.3033 \\
    \midrule
    
    DINO & 3.07 & 0.3616 \\
    \textbf{DINO-SSPS} & \bfseries 2.53 & \bfseries 0.2843 \\
    
    \bottomrule
  \end{tabular}
\end{table}

Table~\ref{tab:eval} presents the final performance of SimCLR and DINO, using ECAPA-TDNN, with and without SSPS (bottom), compared to other state-of-the-art SSL approaches for SV (top). SSPS ($K=25{,}000$ and $M=1$) improves the EER and minDCF of both frameworks on the VoxCeleb1-O benchmark. The best performance is obtained by DINO with SSPS, reaching 2.53\% EER and 0.2843\% minDCF. Additionally, SimCLR provides performance on par with DINO by achieving 2.57\% EER and 0.3033 minDCF, a remarkable improvement over its baseline (58\% relative EER reduction). This finding is very prospective as SimCLR relies on a simpler training framework and reaches the best SSL performance using the Supervised positive sampling strategy in Table~\ref{tab:pos_sampling_bottleneck}, which implies that there is potential for further improvement of SSL contrastive-based methods. Therefore, SSPS improves the SV performance of major SSL frameworks, reducing the performance gap with the fully supervised baseline of 1.34\% EER. Finally, the proposed SimCLR-SSPS and DINO-SSPS outperform other state-of-the-art SSL methods for SV by providing an explicit solution to their main limitation.

\subsection{Visualization of speaker representations}

\begin{figure}[t]
  \centering
  \begin{subfigure}[b]{0.495\linewidth}
    \centering
    \includegraphics[width=\linewidth]{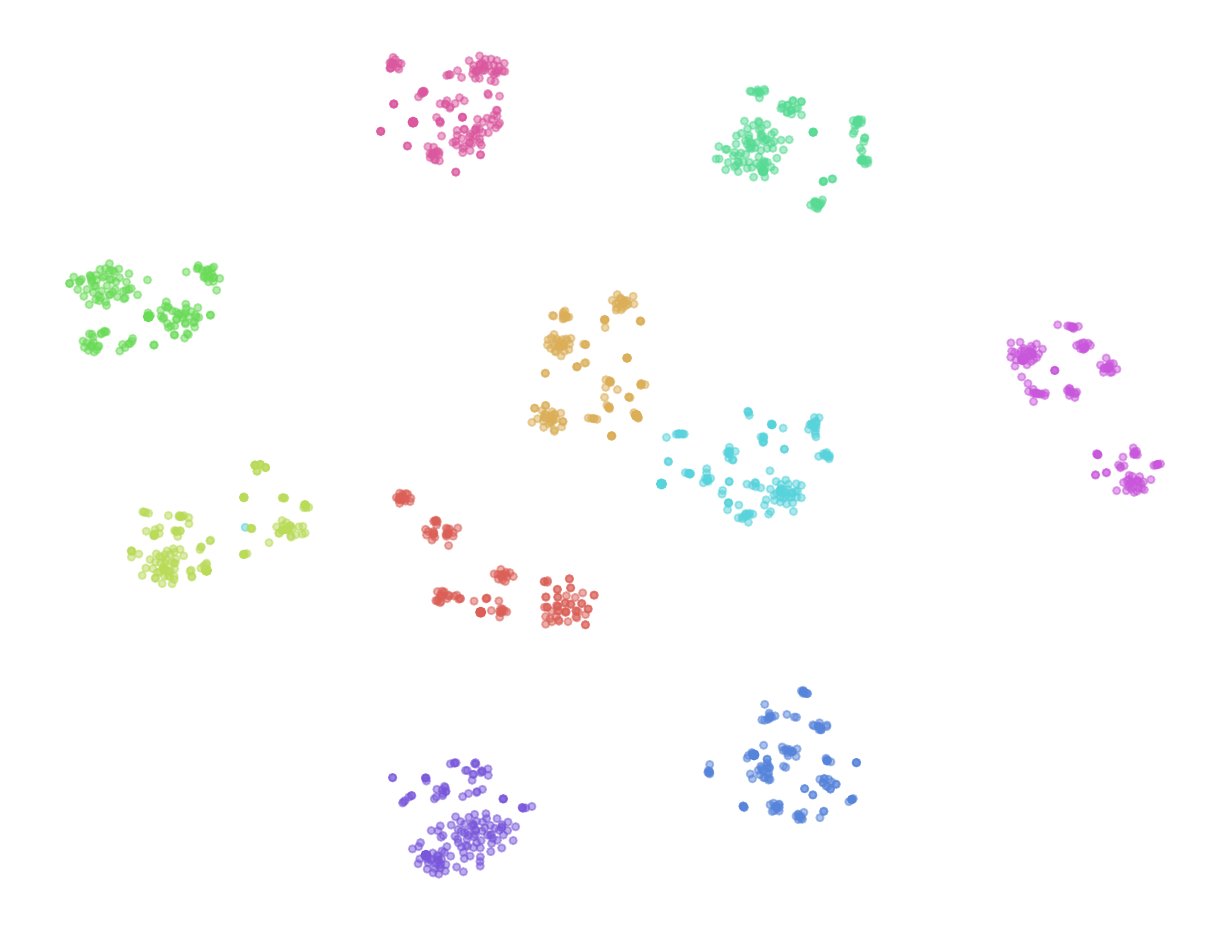}
    \caption{SSL}
  \end{subfigure}
  \hfill
  \begin{subfigure}[b]{0.495\linewidth}
    \centering
    \includegraphics[width=\linewidth]{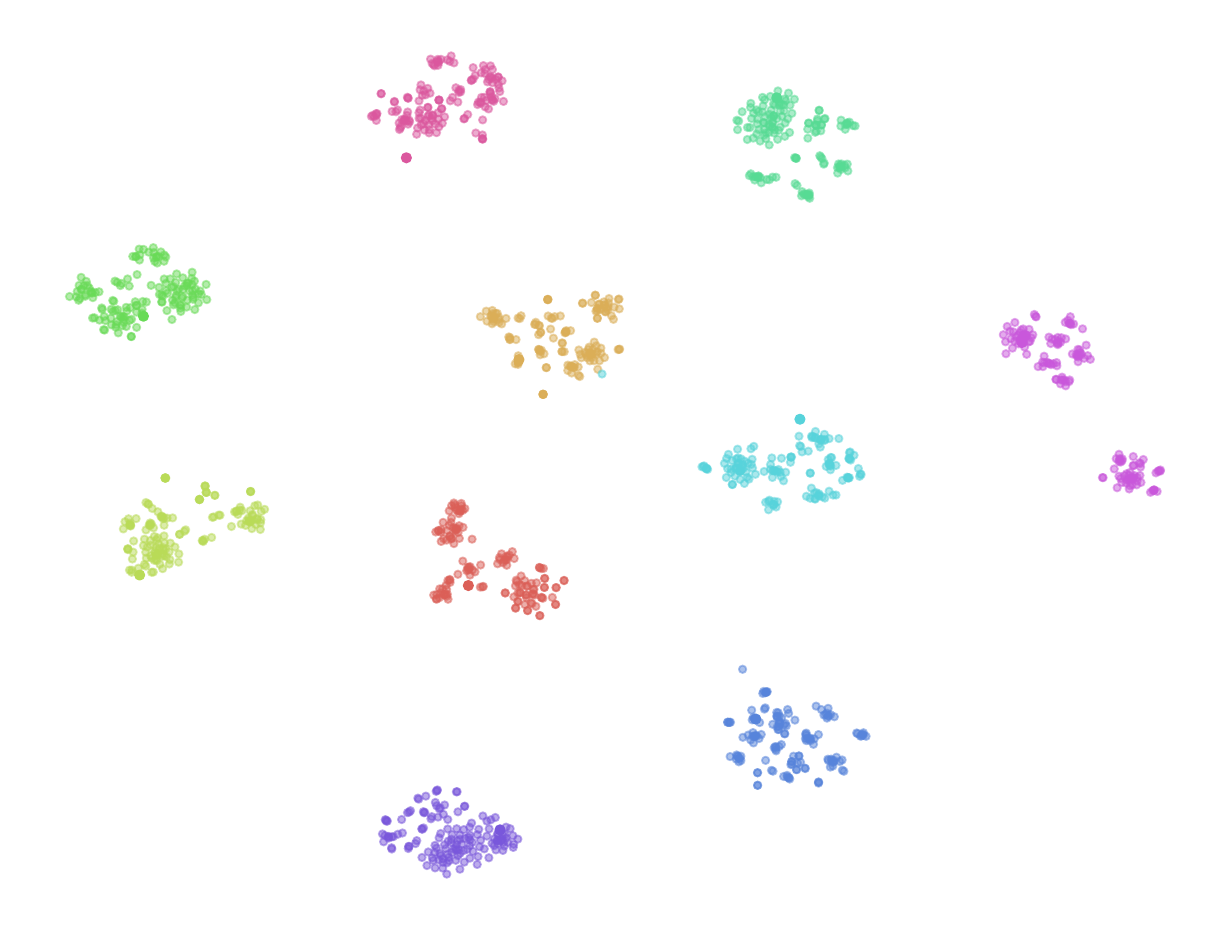}
    \caption{SSPS}
  \end{subfigure}
  \caption{t-SNE of 10 speakers from VoxCeleb1 with SSL and SSPS positive sampling using SimCLR (ECAPA-TDNN).}
  \label{fig:tsne}
\end{figure}

To illustrate the improvement in class compactness, Figure~\ref{fig:tsne} presents the t-SNE of 10 speaker representations from the test set with SSL and SSPS positive sampling techniques. SSPS allows for reducing intra-class variance by matching different recording conditions to the same speaker identity during the training. Note that same-speaker representations already far apart in the latent space are not considered to belong to the same speaker class when employing SSPS, as detailed in~\cite{lepage2025SSLSVBootstrappedPositiveSampling_submitted}.
\section{Conclusions}
\label{sec:conclusions}

This work proposes a new method for sampling positives in SSL frameworks to address the limitations of the same-utterance positive sampling. SSPS samples positives that belong to the same or a neighboring cluster as their corresponding anchor in latent space, such that anchor-positive pairs originate from the same speaker identities but different recordings. This approach achieves SOTA performance with SimCLR and DINO on VoxCeleb1-O by reducing intra-speaker variance.

\section{Acknowledgements}
This work was performed using HPC resources from GENCI-IDRIS (Grant 2023-AD011014623) and has been partially funded by the French National Research Agency (project APATE - ANR-22-CE39-0016-05).

\bibliographystyle{IEEEtran}
\bibliography{biblio,biblio_additional}

\end{document}